\documentclass[reprint,sort&compress,longbibliography,superscriptaddress,amsmath,aps,showpacs,pra]{revtex4-2}

\usepackage{silence}
\usepackage{xr-hyper}

\usepackage{hyperref}
\usepackage{placeins}

\usepackage{multirow}

\usepackage{amssymb}
\usepackage{amsmath}
\usepackage{graphicx,nicefrac}
\usepackage{color}
\hypersetup{
     colorlinks   = true,
     citecolor    = blue
}
\usepackage{physics}
\usepackage{float}
\usepackage{bbold}

\usepackage[justification=justified]{caption}
\usepackage{subcaption}
\usepackage[normalem]{ulem}

\renewcommand{\vec}[1]{\boldsymbol{#1}}
\graphicspath{{Figs/}}
\newcommand{\be}{\begin{equation}}
\newcommand{\ee}{\end{equation}}
\newcommand{\bea}{\begin{eqnarray}}
\newcommand{\eea}{\end{eqnarray}}

\usepackage{natbib}
\begin{document}
\title{Flat bands in chiral multilayer graphene}
\begin{abstract}
 We study the formation and properties of perfectly-flat zero energy bands in a multi-layer graphene systems in the chiral limit. Employing the degrees of freedoms of the multi-layer system, such as relative twist-angle and relative shifts, in a way that preserves a set of symmetries, we define a two-dimensional parameter plane that hosts lines of two and four flat bands. This plane enables adiabatic continuation of multi-layer chiral systems to weakly coupled bi- and tri-layer systems, and through that mapping  provides tools for calculating the Chern numbers of the flat bands. We show that a flat band of Chern number $C$ can be spanned by $C$ effective Landau levels, all experiencing an effective flux of $1/C$ flux quantum per unit cell, and each carrying its own intra-unit-cell wave function. The flat bands do not disperse under the effect of a perpendicular magnetic field, and the gap to the dispersive bands closes when the externally applied flux cancels the $1/C$ effective flux.  
\end{abstract}
\author{Roi Makov}
\affiliation{Department of Condensed Matter Physics, Weizmann Institute of Science Rehovot 7610001, Israel}
\author{Francisco Guinea}
\affiliation{IMDEA Nanociencia, C/ Faraday 9, 28049 Madrid, Spain}
\affiliation{Donostia International Physics Center, Paseo Manuel de Lardiz\'abal 4, 20018, San Sebasti\'an, Spain}
\author{Ady Stern}
\affiliation{Department of Condensed Matter Physics, Weizmann Institute of Science Rehovot 7610001, Israel}

\maketitle

\section{Introduction}
Systems made of twisted graphene monolayers have proven to be a powerful platform to explore strongly correlated phases and topological flat bands\cite{Cetal18,Cetal18b,AM20,BDEY20}, due to the tunability of the Fermi velocity and electronic band structure. While for twisted bi-layer graphene (TBG) the twist angle is the major controlled degree of freedom, for twisted trilayers (TTG) there are two twist angles and one inter-layer shift degree of freedom that may be controlled. These extra degrees of freedom allow tuning of new and interesting electronic structures, including a band structure with four flat bands per valley, non-zero total Chern numbers of the flat bands in each valley and significantly larger gaps around these bands. 

Beyond the three-layer case, twisted multilayer graphene systems (TMG) have significantly more degrees of freedom that may be engineered, such as the twist angles and relative shifts of each layer.  Certain multilayers, defined by twist angles between neighboring layers are such that $\theta_{n,n+1}=-\theta_{n+1,n+2}$, and form the so called symmetric multilayers. These multi-layers may be reduced to superpositions of twisted bilayers and monolayers\cite{KKTV19,Hetal21,Ketal21,Petal22,Zetal22,Betal22,Letal22b}. This simplification cannot be carried out in generic multilayers\cite{WZY23,MGM23,PT23,Detal23,GMM23,Cetal23,NKK23,R23,KLLD23,Uetal23,Yetal23,Fetal24,Hetal24,Petal24,Petal24b} (see also earlier references in\cite{CWG19,MRB19,LWM19,WZS20,ZCMLK20}).

In this work we analyze perfectly-flat bands that occur in TTGs and TMGs at the chiral limit. In the absence of inter-layer tunneling, a stack of $N_L$ graphene layers harbors $N_L$ Dirac cones per valley. In the situations we consider, inter-layer tunneling leaves $n_D$ Dirac cones gapless and unflattened. The rest, $N_L-n_D$ take part in the construction of flat bands. Out of these, $n_A$ construct flat bands on the $A$-sublattice, and $n_B$ on the $B$ sublattice.  When presented in a sub-lattice basis, the flat bands carry non-zero Chern numbers. We analyze how the Chern numbers of the flat bands are related to $n_A,n_B$ and the symmetries of the system, suggesting efficient ways to infer the Chern numbers of the flat bands. We map each band with a Chern number $C$ to a set of $C$ lowest Landau Levels, each carrying one state per $C$ unit cells. The levels have an effective flux that corresponds to $1/C$ flux quanta in each unit cell, and are distinguished by their sub-unit cell wave-functions. When subjected to a perpendicular magnetic field, the flux associated with the external field adds to the effective flux. When the two cancel one another, i.e., when the external flux carries $-1/C$ flux quantum per unit cell, the gap between the zero energy bands and the dispersive bands must close.  

 We connect the possible flat band structures in TMG to those in TBG, emphasizing the role of symmetries in the determination of the conditions needed for the formation of flat bands. We then study the evolution of flat bands when the inter-layer tunneling amplitudes are varied. We find it illuminating to introduce two inter-layer tunneling amplitudes $f_1, f_2$ such that pairs of consecutive layers have different tunnel coupling strength, see Fig. (\ref{fig:PerturbationStructure}) . When $f_1=0$ or $f_2 = 0$ every other pair of layers is decoupled, and the system reduces to decoupled TBG systems and, for an odd number of layers, one monolayer graphene. We show that there are curves in the two-dimensional parameter plane $f_1,f_2$ that preserve the flat bands and their properties, including their number and Chern numbers. In some cases, the flat band curves extend from the decoupled TBG case (either $f_1=0$ or $f_2=0$) to the equal-amplitudes case $f_1=f_2$ and are therefore adiabatically connected to flat bands in TBG. Interestingly, in other cases TMG solutions may not be connected adiabatically to TBGs. For $f_1\approx f_2$ these parameters can be viewed as directly related to the twist angles because to first order in $f_1 - f_2 $ changes in $f_1$ and $f_2$ are equivalent to changes in the angles. The stability we find may be viewed as stability with respect to variation in the twist angles. 

The structure of the paper is as follows: In Sec. (\ref{sec:Model}) we define the Chiral model for the cases we consider, and generalize the method used to find flat bands to the TMG cases. In Sec. (\ref{sec:ChernMag})  we show the analogy between a set of Landau Levels and the perfectly flat bands in graphene. In Sec. (\ref{sec:FlatBandCurves}) we introduce a method to adiabatically connect the perfectly flat bands in TMG to those in TBG and following it show in  Sec. (\ref{sec:GeoFlatBands}) that this connection determines the behavior of these flat bands. We calculate a variety of numerical examples to demonstrate the accuracy and applicability of the method in Sec. (\ref{sec:numerics}). We conclude in Sec. (\ref{sec:conclusions}) with a short summary of our results and a few comments on possible implications. 

\section{Continuum Chiral model for twisted multilayer graphene} \label{sec:Model}
In the following, we consider an $N_L$ layers TMG system with relative twists $\theta_{l,l+1} = \theta_l - \theta_{l+1} = p_{l,l+1}\theta_0 $ for a small $\theta_0\ll 1$ and relative shifts $d_{l,l+1}$ that maintain the $C_3$ symmetry.  In Sec. (\ref{sec:ChernMag}) we limit ourselves to the helical case, where $p_{l,l+1}$ is a constant.
All later sections are valid also for commensurate angles. We neglect a small deviation from the moiré unit cell in consecutive pairs which leads to a meta-moiré. This meta-moiré can be viewed as a position-dependent shift between layers\cite{MGM23}, and the size of the meta-moiré unit cell grows as $( d / \theta )^2$, where $d$ is graphene's lattice constant (see appendix \ref{app:FullModelDerivation}). We assume next that the shift between layers, $d_{l,l+1}$ is constant, which is a reasonable approximation to the local properties of the multilayer when $\theta \ll 1$. Alternatively, a combination of small heterostrains and twists, of order $\theta^2$ can lead to the existence of a single moiré scale of order $\theta$. 

We use the continuum approximation to the interlayer tunneling in order to calculate the electronic structure of the multi-layer\cite{LPN07,BM11}. The existence of magic angles and flat bands in the multilayers is determined using the chiral approximation\cite{SGG12,TKV19}, where hoppings between sites in the same sublattice in neighboring layers are neglected. Then, the Hamiltonian is of the form (see Appendix \ref{app:FullModelDerivation}):
\begin{equation} \label{eq:Ham1}
    \mathcal{H}_{TMG} = \begin{pmatrix} 
     0 & \mathcal{D}^\dagger(r) \\
     \mathcal{D}(r) & 0 
    \end{pmatrix}
\end{equation}
where the $\mathcal{D}^\dagger$ is the operator on sublattice A and the $\mathcal{D}$ is the operator on sublattice B, 
The block matrix $\mathcal{D}$ is defined by the intralayer terms, 
\begin{equation}
     \left({\mathcal{D}}\right)_{l,l} = -2i\overline{\partial} 
\end{equation}
and nearest layer tunneling terms 
\begin{equation} \label{eq:definitionD}
\begin{split}
      \left({\mathcal{D}}\right)&_{l,l+1}=f_{l,l+1}\sum_{l=1}^{N_L-1} U(p_{l,l+1}(r-d_{l,l+1}))  \\
      \left({\mathcal{D}}\right)&_{l+1,l}=f_{l+1,l}\sum_{l=1}^{N_L-1} U(p_{l,l+1}(d_{l,l+1}-r))   \\ 
\end{split} 
\end{equation}
Here $U$ is the standard chiral graphene potential $U(r)=e^{-iq_1 r} + e^{i\phi}e^{-iq_2r}+e^{-i\phi}e^{-iq_3r}$, $\phi = 2\pi/3$, $q_1 = k_\theta(0,-1)$, $q_{2,3}= k_\theta(\pm\sqrt{3}/2,1/2)$ and $k_\theta=2K_D\sin(\theta/2)$. 

We now define the tunneling  operator $\mathcal{P}$ such that,
\begin{equation} \label{eq:tunpot}
    \mathcal{D}= -2i \overline{\partial} I + \mathcal{P}
\end{equation}

For TBG and TTG it was shown that zero-energy flat bands are present when a point $r_0$ exists in which zero-energy states at Dirac points $\Gamma, K, K^\prime$ are linearly dependent, i.e. the  zero-energy spinors satisfy $\sum_i c_i \Psi_{K_i}(r_0)=0$, where $c_i$ are constants. The search for such an $r_0 $ starts by searching for a zero of the Wronskian, $W$ (defined below), and then using that $W(r) = 0 $ to construct linearly dependent Dirac spinors. The extension of these spinors to a full flat band makes use of Jacobi theta functions \cite{GMM23}\cite{guerci2023nature} \cite{PT23}. Details are reviewed in Appendix \ref{app:WronskianFlatBands}. 

We generalize this approach to analyze a general TMG system. The Wronskian is, 
\begin{equation}
    W(r) = \det \left(\Psi_{K_1} \dots \Psi_{K_{N_L}}\right)
\end{equation}
where $\Psi_{K_i}({\vec r})$ are the Dirac zero modes such that $\mathcal{D}\Psi_{K_i}=0$. In the small angle limit, these modes occur at the $K,K',\Gamma$-points of the Brillouin zone formed in the twisted system. Using the Jacobi formula and Liouville's theorem we can see that the Wronskian is position independent $W(r)= W(0)$ and there are flat bands if and only if the Wronskian is zero. When $W=0$ only  $m < N_L$ independent Dirac zero modes exist and every possible zero energy solution can be expressed using these Dirac zero modes,
\begin{equation}
   \psi(r) =  \sum_{i=1}^m c_i(r) \Psi_{K_i}(r)
\end{equation}
where $\psi(r)$ is a zero energy solution. When $W = 0$ we will necessarily have a point $r_0$ in which the $m$ Dirac zero modes are linearly dependent. For each such point $r_0$ we can create a perfectly flat band by properly choosing $c_i(r)$ such that the states satisfy the Bloch form for all points in $K$ space. Details of this generalization are given in Appendix \ref{app:WronskianFlatBands}. 

\section{Chern numbers}\label{sec:ChernMag}

The Chern numbers, $C$, of the flat bands are determined by the dimension of the solution space, which is number of independent Dirac zero modes from which we construct our flat band. Generally, we show in Appendix (\ref{app:ChernNumbers}) that
\begin{equation} \label{eq:ChernEquality}
   N_L =  n_D +  n_{A}+ n_{B} 
\end{equation}
where $N_L$ is the number of layers, $n_D$ is the number of Dirac cones for which the Dirac velocity is not made to vanish by the periodic potential and  $n_{A},n_{B}$ are the number of Dirac zero modes used to construct flat bands on sublattices $A$ and $B$ respectively.

The Chern numbers are constrained by $n_A$ and $n_B$ such that flat bands constructed from a single vanishing point $r_0$ will have Chern numbers $|C| = n_{A,B} $. We will show in Sec. (\ref{sec:GeoFlatBands}) that the numbers $n_A$ and $n_B$ can often be determined by symmetries and by adiabatically connecting the system to a set of decoupled TBGs. 


The $C = 1$ flat bands in chiral TBG were shown to be spanned by a basis of functions 
\begin{equation}
    \psi_m(\bf{r})=G({\bf r})(z-z_0)^me^{\frac{2\pi}{C A_m}\left|z-z_0\right|^2}
    \label{LLLG1}
\end{equation} where $A_m$ is the area of the unit cell, $z_0$ is the complex representation of $r_0$, and $G({\bf r})$ is periodic up to a phase. 
\cite{AdyYardenMag}. As we now show, flat bands with $C=N$ are spanned by $N$ copies of the basis (\ref{LLLG1}), where each of the copies  has its own periodic function $G$, and corresponds to a lowest Landau levels with a flux of $1/N$ flux quanta per unit cell. Consequently, each copy has $1/N$ of the density of states of the full flat band. 

\begin{figure}[t]
\vspace{0.5cm}
   \includegraphics[width=0.45\textwidth]{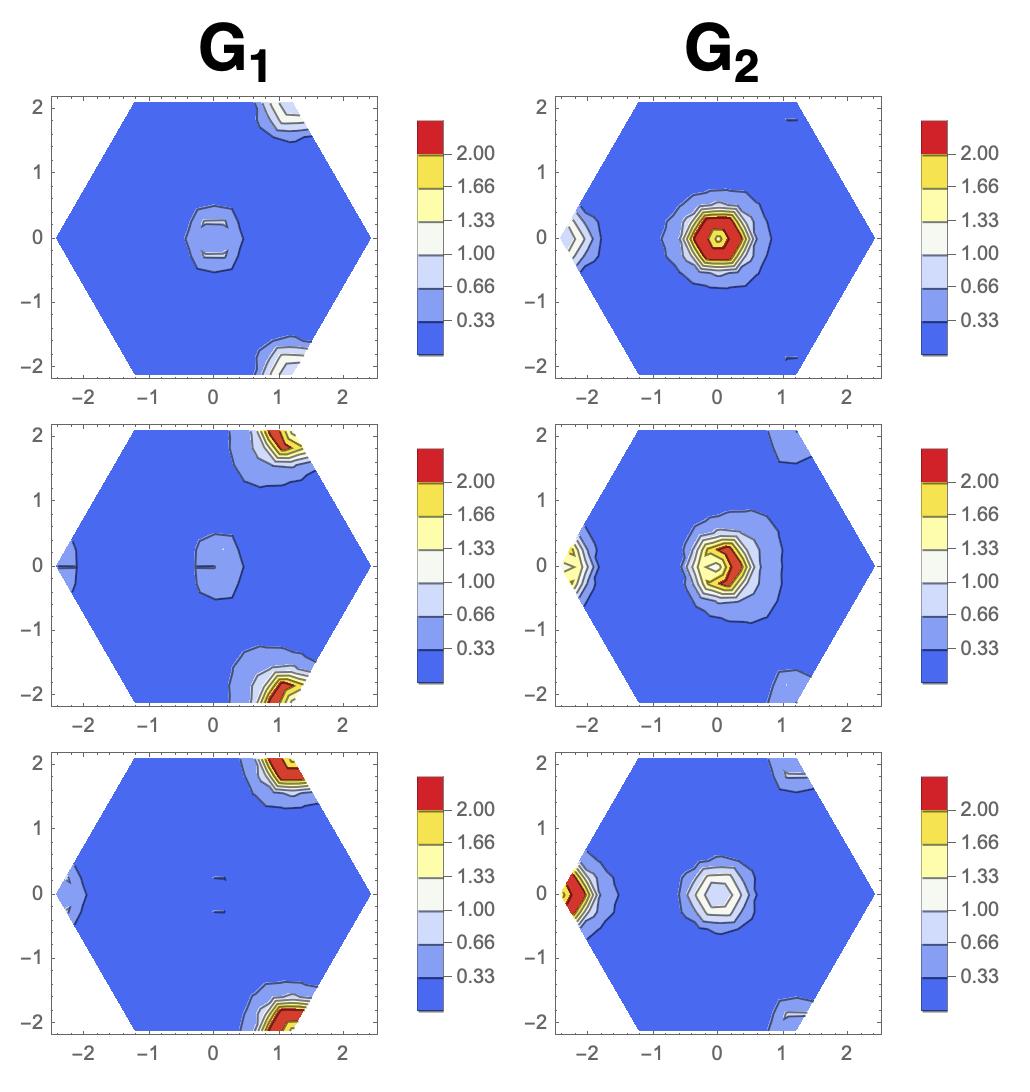} 
   \caption{The  periodic functions $|G_1|^2$ and $|G_2|^2$ as a function of position within the unit cell, for each layer for a tri-layer in the ABA case. The color coding is in arbitrary units.}
 \label{fig:GFuncs}
\end{figure}

The construction of a $C=N$ flat band requires the number of independent Dirac zero modes to be $m=N$ \cite{PT23}. When that happens we can take these $m$ Dirac zero modes and construct $N$ orthogonal flat bands with $C=1$ each, $\{\Psi_i\}_{i=1}^N$ , with the unit cell of each flat band increased by a factor of $N$ relative to the moire unit cell. 
\begin{equation} \label{eq:LandauLevels}
    \Psi_{i,m}(\bf{r}) = \left(z-z_0\right)^m e^{-\frac{2\pi}{C A_m}\left|z-z_0\right|^2}G_i(\bf{r})
\end{equation}
where $G_i(r)$ are periodic functions, and the wave-functions $\Psi_i(r)$ are normalizable, see Appendix \ref{app:LandauLevelComparison}. We plot the functions  $|G_1|^2$ and $|G_2|^2$  for a tri-layer in the ABA case in Fig. (\ref{fig:GFuncs}). We can see that these functions are concentrated  in different regions of the unit cell.   The mapping of each of the $N$ bands to a Landau level implies that the Chern number of each of the bands is $C=1$. 

Introducing a magnetic field, $b$, perpendicular to the TMG will change the number of states, $n_i(b)$, in each Chern band, as this number changes with $b$ according to $n_i(b) = n_0 \left(1 +  \frac{\Phi}{C\Phi_0}\right)$, where $\Phi$ is the flux per unit cell and $\Phi_0$ is the flux quantum. When $n_i(b) \le 0$ the Chern number of the bands must change as one of the bands empties and the gap to the dispersive bands must close \cite{AdyYardenMag}. We also consider the same question from a Landau level perspective by defining the Hamiltonian with magnetic field as $\mathcal{D}_b = \mathcal{D_0} + \frac{1}{2}I zb $. The new Hamiltonian has solutions $\Psi_i^b(r) = e^{-\frac{1}{4}b|z-z_0|^2}\Psi_i^0(r)$ which generates new normalizable wave-functions,
\begin{equation}
    \Psi_{i,m}^b(\bf{r}) = \left(z-z_0\right)^m  e^{-\left(\frac{b}{4}+\frac{2\pi}{C A_m}\right)|z-z_0|^2} G_i(\bf{r})
\end{equation}
This mapping allows us to see more directly that the magnetic field introduces a topological phase transition at $-\phi_0/C$, which is the point where $\frac{n}{4}+\frac{2\pi}{NA_m}=0$ and the wave functions are no longer normalizable. This  is the same condition as $n_i(b)=0$.  The result was previously observed for chiral TBG \cite{AdyYardenMag}. However, for TBG the topological phase transition does not exist away from the chiral limit, since the total Chern number of the flat bands is zero. While the sub-lattice separation between the flat bands is not maintained for non-chiral TMG, the total Chern number is conserved. As such, whenever the total Chern number per valley is non-zero the system will have Chern bands even away from the chiral limit. That means a topological phase transition will happen in a variety of TMG cases, including two flat bands in TTG and twisted five layer graphene.

\section{Flat bands in Twisted Multilayer Graphene} \label{sec:FlatBandCurves}
We now study the case in which the interlayer tunneling amplitudes $\{f_l\}$ can only take two possible values $f_1$ and $f_2$.  We will show that there exist curves in the parameter space $f_1$, $f_2$ along which there are flat bands. Since for $f_1\approx f_2$ these parameters can be viewed as directly related to the twist angles, this finding implies that there are curves in twist angles parameter plane along which there are flat bands\cite{PT23,Fetal24}. 

\begin{figure}
    \includegraphics[width=0.5\textwidth]{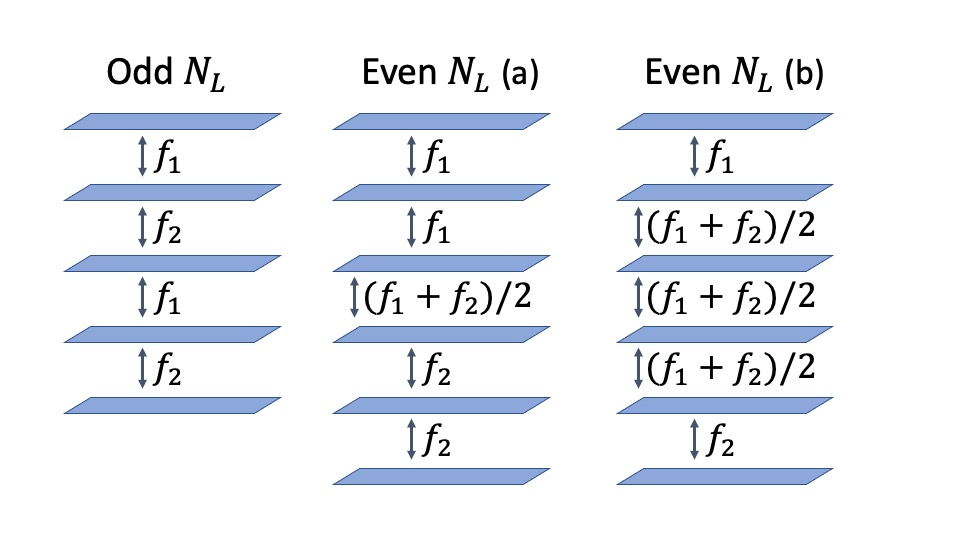}    
    \caption{The multi-layer systems we consider, where $f_{1,2}$ are the tunnel coupling parameters. The configurations of perturbation parameters $f_1$ and $f_2$ are used to connect adiabatically between TMG and simpler cases. For an odd $N_L$ we always start from decoupled TBG at $f_{1}=0$ or $f_2 =0$. For an even $N_L$ we consider two separate configurations (a) and (b) that  connect adiabtically to a pair of TMG with $(N_L -1)/2$ layers and a pair of opposite twist TBGs respectively.}
    \label{fig:PerturbationStructure}
\end{figure}

For most of our discussion we assign the values $f_1$ and $f_2$ as alternating values between consecutive pairs of layers, but a few other configurations will be used as well, see Figure \ref{fig:PerturbationStructure}.

We can write the Hamiltonian in terms  of the parameters $f_1$ and $f_2$ as,

\begin{equation}
    \mathcal{D} = -2i\overline{\partial} I_{n\times n} + f_1\mathcal{P}_1 + f_2 \mathcal{P}_2
\end{equation}
where the operator $\mathcal{P}_1$ and $\mathcal{P}_2$ are the respective operators such that $\mathcal{P} = f_1\mathcal{P}_1 + f_2 \mathcal{P}_2$. 

We now assume that there is a point $f_{1,0},f_{2,0}$ at which perfectly flat bands exist, and use degenerate perturbation theory to analyze small deviations $\delta f_1,\delta f_2$ from this point. At the original point, for every momentum $\bf k$ the zero modes span a subspace whose dimension is the number of flat bands, two or four. When the projections of $\mathcal{P}_1$ and $\mathcal{P}_2$ onto that subspace commute with one another, a deviation $\delta f_1$ may be matched with a corresponding deviation  $\delta f_2$ such that at least two eigenvalues remain zero. 
As we discuss below, 
certain symmetries need to exist for these two projected operators to commute. 
Importantly, 
as shown in Sec \ref{sec:Model} we do not need to analyze every momentum $\bf k$, since we can construct an entire flat bands from a single zero mode. Specifically, we can choose that mode to be 
at the $\Gamma$-point, which allows symmetries that are not local in $K$ space to constrain the operators $\mathcal{P}_1$ and $\mathcal{P}_2$. 

We will examine three key cases, distinguished by the symmetries that they preserve. In order to extend the TBG symmetries to a multi-layer case, we define extended layer-space $N_L\times N_L$ Pauli matrices  as,
\begin{align} \label{eq:extPauli}
    \begin{split}
        (\eta_x)_{kj} &= \delta_{k+j=N_L+1}\\
        (\eta_y)_{kj} &= (-i)^{k+1} \delta_{k+j=N_L+1}\\
        (\eta_z)_{kj} &= (-1)^{k+1}\delta_{k=j}\\
    \end{split}
\end{align}
We now define unitary particle-hole-type symmetry, unitary inversion symmetry and a anti-unitary time-reversal symmetry,
\begin{subequations}
\label{eq:SymSymmetricCase}
\begin{align}
    P: & \eta_y (r\to -r)(f_1 \leftrightarrow f_2) \label{eq:PSym} \\ 
    C_{2y}: &\eta_x (x\to -x) (f_1 \leftrightarrow f_2) \label{eq:CYTSym} \\
    C_{2z}: & \sigma_x (r\to -r) \\
    T: & K  \\
    C : & \sigma_z
\end{align}
\end{subequations}
where $\eta_{x,y}$ are the extended Pauli matrices (\ref{eq:extPauli}) acting on the layer indices and $\sigma_x$ are regular Pauli matrices acting on the sublattice indices.  The first case is the general case, which preserves only the chiral symmetry $C$  and a $PC_{2y}T$ symmetry.  All configuration described above maintain these symmetries. The second case is the central symmetric case that has a symmetry around the central layer. For this case $\theta_i=\theta_{n+1-i}$ and $d_i =d_{n+1-i}$, and two further symmetries are satified $C_{2y}T$ and $P$. (see Eq. [\ref{eq:SymSymmetricCase}] below). The third case is the unshifted case. It is central symmetric case without shifts,  such that $d_i=0$ and also satisfies the symmetry $C_{2z}T$.

\subsection{Flat band curves of two flat bands}
At a point in the parameter space ($f_1$, $f_2$) in which there are two flat bands, 
the two-band subspace may be represented by a single set of Pauli matrices. 
The projected operators at the $\Gamma$ point then satisfy,
\begin{equation}
    P_{i} = \sum_j g_{ji}s_j
\end{equation}
where $s_j$ are the Pauli matrices within the projection space. We can now use symmetries to significantly simplify these representations. The coefficients $g_{ij}$ would in general depend on $k$, however as we are looking at the $\Gamma$ point we can treat them as scalar coefficients.  We represent the chiral symmetry as 
$s_z$, which limits the sum over $j$ to $x$ and $y$. 

Furthermore, we can use the particle-hole time reversal symmetry $PC_{2y}T$ represented as $\eta_z K (y \to -y)$, where $\eta_z$ is an extended Pauli matrix for the layers. This operator  anti-commutes with the Hamiltonian. In the flat band subspace it is $s_0 K$, such as to maintain correct commutation relations with the Chiral symmetry and the Hamiltonian. Then the projected $\mathcal{P}_i$ operators can only be represented as,
\begin{equation}
    \mathcal{P}_{i} = g_{i}s_y
\end{equation}
and therefore $\left[\mathcal{P}_1,\mathcal{P}_2\right] = 0$ within the projection space. Then, for any shift $\pm \delta f_1$ there  is as shift $\pm \delta f_2$ that maintains two degenerate flat bands. Consequently, with these symmetries and two flat bands we are guaranteed to have continuous flat band curves in parameter space $f_1$, $f_2$. 

\subsection{Flat band curves of four flat bands}

When we have four flat bands at a point $f_1$, $f_2$ we can represent the projected operators using two Pauli matrices $s_i$ and $\tau_j$. 
The projected operators $\mathcal{P}_1, \mathcal{P}_2$ at the $\Gamma$ point, 

\begin{equation}
    \mathcal{P}_{i} = \sum_{j,k} g_{kji}\tau_js_{k} 
\end{equation}
where, once again $g_{kji}$ are in general a function of momentum but  can be treated  as constants at the $\Gamma$ point. We can now use both earlier symmetries, but this time choosing to represent $\eta_z K(y\to-y) $ as $\tau_zK$ giving the following reduced representation,

\begin{align}
    \begin{split}
        \mathcal{P}_{i} &=  g_{xxi}\tau_xs_x + \sum_{j\in\{y,z\}} g_{yji}\tau_{j}s_y
    \end{split}
\end{align}

The operators $\mathcal{P}_1$ and $\mathcal{P}_2$  are not necessarily commuting and do not guarantee the existence of flat band curves without additional symmetries. For the general case these are the only symmetries, and therefore there will be no curves of four flat bands in that case. Points of four flat bands are still possible in the parameter space as the intersection of two different two flat bands curves. 

For the central symmetric case we have two additional symmetries: the unitary symmetry $P$   and the anti-unitary symmetry $C_{2y}T$ as defined in eq. (\ref{eq:SymSymmetricCase}). These two symmetries impose new constraints on the representation of $\mathcal{P}$ which allows us to reduce the constants $g_{xxi}, g_{yyi}, g_{zyi}$ to a single constant $c$,
\begin{subequations}
\label{eq:SymApplied}
\begin{align}
    f_1\mathcal{P}_1 + f_2\mathcal{P}_2 &= c\left( \tau_xs_x(f_1-f_2) + (f_1+f_2)(\tau_zs_y+\tau_ys_y)\right)\label{eq:firstSym} \\ 
    f_1\mathcal{P}_1 + f_2\mathcal{P}_2 &= c\left( (f_1+f_2)\tau_ys_y \right)\label{eq:secondSym}
\end{align}
\end{subequations}
where Eq. (\ref{eq:firstSym}) shows the application of the first symmetry and eq. (\ref{eq:secondSym}) adds the application of the second symmetry. This result shows that the central symmetric case will have four flat bands curves in contrast to the more general case. That implies that this symmetry might be necessary to get a second order zero of the wave function which is used to construct four flat bands in TTG. 
\\

However, this would not be possible for the unshifted case due to the extra $C_{2z}T$ symmetry prohibiting the reduced operators $\mathcal{P}_1,\mathcal{P}_2$  from having non-zero matrix elements in the subspace of the four bands. Consequently, no four flat bands curves may occur for the unshifted case.

\begin{figure*}
   \includegraphics[width=0.8\textwidth]{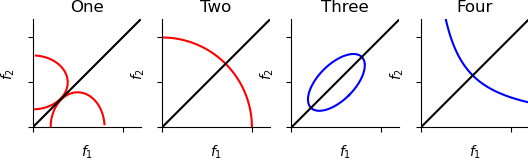} 
   \caption{The four types of possible flat band curves. In red are curves along which two flat bands exist. In blue are curves that preserve four flat bands. Type one is seen in symmetric unshifted models and has four flat bands only at the exact point $f_1=f_2$ where the two lines meet. Type two is the general two flat bands case, seen for all constructions we consider, except the unshifted case with an odd number of layers. Types three and four are four flat band curves seen in the central symmetric models with shifts that preserve the symmetry around the central layer and the $C_3$ symmetry, }
 \label{fig:SampleCurves}
\end{figure*}

\section{Geometric Structure of the flat band curves} \label{sec:GeoFlatBands}
For this section we study the odd $N_L$ configuration, the geometric structures of the flat band curves in the $(f_1,f_2)$ plane for the central symmetric case and the unshifted case are determined by two geometric constraints. The first constraint states that the flat band curve must be orthogonal or tangential to the line $f_1=f_2$. On the line $f_1 = f_2$, the anti-unitary symmetry $C_{2y}T$ exchanges $f_1$ and $f_2$ as defined in eq. (\ref{eq:CYTSym}). Therefore, the small perturbations $\delta f_1$ and $\delta f_2 $ must be interchangeable.  If the perturbations $\delta f_1,\delta f_2$ affect the energy of the states in the band at first order, then the bands will remain flat only when $\delta f_1 = -\delta f_2$. Then, the flat band curve is orthogonal to the line $f_1 = f_2$. If the correction is only of second order, the flat band curve is tangential to the line $f_1=f_2$.   In particular, we can see that four flat band curves would be orthogonal to the line $f_1 = f_2$, as seen from eq. (\ref{eq:secondSym}) which shows the order of the perturbation. 

The second geometric constraint is that a flat band curve for which there is a tangential line that crosses the origin (($f_1=f_2=0$) point) is a four-band curve. 
To see this, let us define such a line from the origin by defining a new parameter $\alpha$ such that the line is 
\begin{align}
    f_2 = \tilde{f}_2 \alpha \;\;\;\;\;
    f_1 = \tilde{f}_1 \alpha 
\end{align}
where $\tilde{f}_2, \tilde{f}_1 $ is the  point in which the line touches the flat band curve. We can now examine a new perturbation in terms of $\alpha$, 

\begin{equation}
    (-2i\overline{\partial})I_{n\times n} + \mathcal{P} + \delta \alpha \mathcal{P} 
\end{equation}

 At the point $(\tilde{f}_1, \tilde{f}_2)$ the Dirac zero mode must vanish at point $r_0$, $\Psi_\Gamma(r_0)=0$, since flat bands exist. Then, being tangential to the flat bands curve implies that the first order correction to the wave function at $r_0$ due to the perturbation $\delta \alpha$ must be zero. This is only possible if the expansion of the zero energy Dirac spinor  $\Psi_\Gamma$ to first order in the deviation from $r_0$ does not depend on $\overline{z}$ such that $\mathcal{P}\Psi_\Gamma=0$ near $r_0$.  Expanding the $\delta\alpha=0$ wave function to lowest order in $z$ around $r_0$,

\begin{equation}
    \Psi_\Gamma(r+r_0)=z^m \vec{\eta}
\end{equation}
where $\vec{\eta}$ is a vector of the dimension of the number of layers that does not depend on $z$, $z$ is the deviation from $r_0$, and $m$ is an integer.  Due to the symmetry around the central layer, We can now apply the product $C_{2y}T$   to get:
\begin{equation}
    z^m \vec{\eta} = (-z)^m \left(\eta_x \vec{\eta}\right)^*
\end{equation}
This leads to the constraint $m\ge2$, which guarantees a second order zero at that point and therefore four flat bands.

These two geometric constraints mean that we can only have four possible types of flat band curves, shown in Fig \ref{fig:SampleCurves}.  Type one are two-band lines that connect to flat bands of decoupled TBGs, but approach the $f_1=f_2$ line tangentially, and form a four-band state at the point of touch. Type two are two-band curves, which connect to flat bands of decoupled TBGs, cross the $f_1=f_2$ line at a direct angle, and do not have a tangent that crosses the origin. Type three are four-bands curves that do not connect to any decoupled limit, but rather form a closed curve. Type four are four bands curves that approach asymptotically to the $f_1=0$ and $f_2=0$ lines. 
A four flat bands line may never reach either axis and will have two lines tangential to the origin forcing it to be either of type 3 or type 4. Furthermore, on the line $f_1=f_2$ we do not expect to see more than four flat bands because two separate  flat band curves would have to meet there which would not expect without a further symmetry or tuning parameter.

We can use these geometric conditions along with the result from Sec. (\ref{sec:ChernMag}) to calculate the Chern numbers for a variety of cases. If we know $N_L$ and $n_D$ (whose calculation requires only the spectrum close to the Dirac points) we can see from eq. (\ref{eq:ChernEquality}) what would be the sum of $n_A$ and $n_B$. We can then gain further information by considering the $f_1-f_2$-plane curves. For curves of type one and two we can count the number of Dirac spinors $n_A$ when the system is close to the $f_1=0$ or $f_2=0$ axes, and evaluate the Chern numbers of the two bands. 

For curves of type $3$ and $4$ we can only calculate $n_A$ and $n_B$ if $n_A + n_B \le 4$, cases in which there is only a limited number of combinations. 
Furthermore, choosing a different configuration of $f_1$ and $f_2$ could provide more information, e.g. by decomposing the systems to a set of building blocks that includes a trilayer it is possible to directly connect a four flat band solution to the axis. 

The analytical results for low Chern numbers are summarized in Table (\ref{tab:ChernNumbers}). We calculate  numerically the parameters needed to use Table (\ref{tab:ChernNumbers}) for $3\le N_L\le 9$ and present them in Table (\ref{tab:multilayernumerics}).

\section{Example Cases} \label{sec:numerics}

\begin{figure}[t]
   \includegraphics[width=0.3\textwidth]{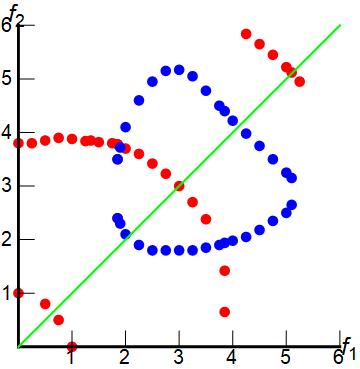} 
   \caption{Numerical calculation of the perturbation for an ABA tri-layer, showing type two flat band curves  for two flat bands in red and type three for four flat bands in blue}.
 \label{fig:ABA}
\end{figure}

We now consider numerically several example cases for constructions that already have known solutions on the line $f_1=f_2$. We consider the central symmetric trilayer case and the unshifted trilayer case. In the central symmetric case, we could theoretically have flat band curves of types 2,3 and 4 but specifically in the trilayer case we only observe types 2 and 3 as can be seen numerically in Fig \ref{fig:ABA}. 

\begin{figure}[t]
   \includegraphics[width=0.3\textwidth]{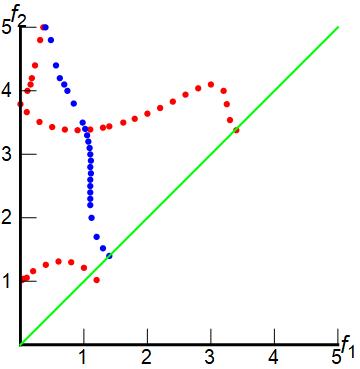} 
   \caption{As in Fig.[\ref{fig:ABA}], for a five layer stack.}
 \label{fig:ABA5}
\end{figure}

\begin{figure}[t]
   \includegraphics[width=0.3\textwidth]{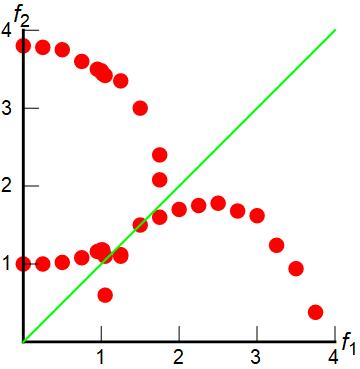} 
   \caption{Numerical calculation of the flat bands for an AAA tri-layer, showing type 1 flat band curves  for two flat bands resulting in a four flat band solutions at a single point}.
 \label{fig:AAA}
\end{figure}

In the unshifted case we can not preserve a four-band curve and as such we can only have flat band curves of types one and two.  This case has an extra time-reversal-like symmetry $PCC_{2z}T$, which is an anti-unitary symmetry that squares to $-1$ for an odd number of layers and to $+1$ for an even number of layers. From Kramer's theorem we can see that for an odd number of layers this symmetry must generate a new pair of flat bands at $f_1=f_2$. As such only type one curves are possible, and on the line $f_1=f_2$ there must be two independent pairs of flat bands. This can be seen numerically for three layers in Fig \ref{fig:AAA}.

\section{Conclusions}\label{sec:conclusions}

In the last few years, studies of twisted van der Waals systems have re-emphasized how the combination of narrow Bloch bands and electron-electron interaction leads to the realization of sophisticated many-body systems. As a consequence, much theoretical and experimental effort is presently devoted to the understanding of circumstances in which narrow electronic bands occur. 

In this paper we studied the formation of perfectly flat bands in the chiral limit of twisted $N_L$-layer graphene systems. While  $N_L$ decoupled layers carry $N_L$ Dirac cones per valley per spin, a proper combination of inter-layer tunneling magnitude, twist angles and shifts leads to the flattening of some of these cones into perfectly flat bands.  Here, we constructed a method for creating such bands and analyzed their properties. While for bi-layer graphene perfectly flat bands occur at a discrete set of magic angles, we showed that for multi-layer systems flat bands occur along lines in the two-parameter plane of alternating tunneling amplitude or alternating twist angles. The flat bands carry Chern numbers, and we showed how their Chern numbers are determined by general principles, in particular the system's symmetries. We then showed how the flat bands may be spanned by a basis that is very close to that of a lowest Landau level, being made of a product of a lowest Landau level wave-function multiplied by an intra-unit-cell wave function. A perfectly flat band of Chern number $C$ may be mapped onto a set of $C$ lowest Landau levels with an effective flux of $1/C$ flux quanta per unit cell; with $1/C$ states per unit cell per each of the $C$ levels;  and with a different intra-unit-cell wave functions distinguishing between the $C$ levels. 

When subjected to an external perpendicular magnetic field the flat bands remain flat, but the number of states within each band changes. When a band is emptied, the gap between the flat and the dispersive bands must close. This happens when the external flux cancels the  effective flux of $1/C$ flux quanta per unit cell. When the total Chern number per valley is non-zero, such gap-closing takes place even when the system deviates from the chiral limit. 

The mapping of a flat band with $C>1$ to a set of lowest Landau levels that differ in their intra-unit-cell distribution provides a way to think about many-body fractional Chern insulator states in such bands. In particular, it allows for a mapping of these bands onto a lowest Landau level with $C$ species, with the short range part of electron-electron interaction becoming a matrix in the space of species. Similar considerations occur in multi-component fractional quantum Hall states \cite{SRBR2013}. 


Our analysis of the evolution of flat bands curves in the $f_1-f_2$  plane suggests the adiabatic continuation of flat bands from the symmetric $f_1=f_2$ line to the decoupled $f_1=0$ or $f_2=0$ lines. This continuation provides a way to analyze flat bands in $N_L$-layer TMG, including their Chern numbers, through their reduction to a set of decoupled systems with a small number of layers. For example, a tri-layer  is fully gapped ($n_D=0$) at $f_1=f_2$, while at the $f_1=0$ line it is reduced to a gapped bi-layer and a gapless monolayer. The bi-layer has two flat bands with $C=\pm 1$. As $f_1$ is turned on the gapping of the Dirac cone in the monolayer changes $n_D$ from one to zero, and changes the Chern numbers of the flat bands to $2,-1$ or $1,-2$. 
Generally, we note that the configurations introduced in Fig. (\ref{fig:PerturbationStructure}) do not exhaust the possible ways for such reductions, and further configurations may be introduced and analyzed for large $N_L$'s. 

Experimentally, variations of the twist angles are routinely implemented at the growth level, and may even be implemented continuously using a Quantum Twisting Microscope \cite{IOABI23}. Variations of the tunneling strength between layers may be implemented by growing of MLGs in which some of the levels are separated by monolayers of insulating hBN. 

\section{Acknowledgements}
We thank Anushree Datta, Christoph Mora and Yarden Sheffer for instructive discussions.. AS thanks the Donostia International Physics Center (DIPC), where this work was started, for hospitality. AS was supported by grants from the ERC under the European Union’s Horizon 2020 research and innovation programme (Grant Agreements LEGOTOP No. 788715 ), the DFG (CRC/Transregio 183, EI 519/71), and by the ISF Quantum Science ands Technology (2074/19). FG acknowledge support from the Severo Ochoa programme for centres of excellence in R\&D (CEX2020-001039-S / AEI / 10.13039/501100011033, Ministerio de Ciencia e Innovaci\'on, Spain);  from the grant (MAD2D-CM)-MRR MATERIALES AVANZADOS-IMDEA-NC, NOVMOMAT, Grant PID2022-142162NB-I00 funded by MCIN/AEI/ 10.13039/501100011033.

\appendix
\section{Full Model derivation} \label{app:FullModelDerivation}
We  derive the chiral continuum Hamiltonian in eq. (\ref{eq:Ham1}) for $N_L$ stacked layers of graphene where each layer $l$ is twisted by angle $\theta_l$, with a relative shift between layers of  $d_{l,l+1}$ and tunneling is only permitted to the nearest layer. The continuum model can be written as a sum of $N_L$ isolated graphene Dirac Hamiltonians and nearest layer tunnelling terms \cite{PT23}\cite{KKTV19} \cite{BM11},

\begin{align}
    \begin{split}
     \mathcal{H} &= - i v_f \sum_{l=1}^{N_L}  \hat{e}_l\hat{e}_l\sigma_{\theta_l} \nabla \\ &+ \sum_{l=1}^{N_L-1} T^{l, l+1}(r-d_{l,l+1})    \\
     &+ \sum_{l=1}^{N_L-1}  T^{\dagger \; l+1, l}(r-d_{l+1,l})  
     \end{split}
\end{align}

Where, $\sigma_{\theta_l}\equiv e^{-i\frac{\sigma_z}{2}\theta_l}\sigma e^{i\frac{\sigma_z}{2}\theta_l}$, $d_{l,l+1}$ is the moire pattern displacement vector with a possible shift. The moire potential $T^{l,l+1}$ between adjacent layers is defined,
\begin{align}
    \begin{split}
        T&^{l,l+1} = \sum_{n=1}^{3} \left(\omega^{l,l+1}_{AA}\sigma_0 +  \omega^{l,l+1}_{AB}Q_n \right) e^{-iq_n^{l,l+1}r} 
     \end{split}
\end{align}
where, $Q_n  =  \left( \sigma_x \cos(n\phi)+\sigma_y \sin(n\phi) \right)$, $\omega^{l,l+1}_{AA}$ and $\omega^{l,l+1}_{AB}$ are the interaction strengths between site $A$ on layer $l$ to sites $A$ and $B$ on layer $l+1$ respectively. $q_n^{l,l+1}$ are the reciprocal lattice vectors for the Moir\'e of each pair of layers. We can define this by setting the first reciprocal vector to be $\hat{y}$ and achieve the other vectors in the lattice by applying a rotation operator $R_\phi$ where $\phi = \frac{2\pi}{3}$. We can now set the three reciprocal vectors for each pairs of layers as,
\begin{equation}
    q_n^{l,l+1}= - R_\phi^n 2K_D\sin\left(\frac{\theta_{l} - \theta_{l+1}}{2}\right) R_{\frac{\theta_l+\theta_{l+1}}{2}} \hat{y}
\end{equation}

We take the chiral limit where $\omega_{AA} = 0 $ and the small angle approximation to neglect the phase factors \cite{PT23} \cite{Fetal24},

\begin{equation}
     q_n^{l,l+1}= - R_\phi^n 2K_D\left(\frac{\theta_{l} - \theta_{l+1}}{2}\right)  \hat{y}
\end{equation} 
As  such the the inter-layer potential is now ,
\begin{align}
    T^{l,l+1}(r)&= \omega^{l,l+1}_{AB} \sum_{n=1}^{3} Q_n e^{-iq_n^{l,l+1}r}
\end{align}
 
 We define the interlayer tunneling coupling parameters $f_{l,l+1}$ by $\omega^{l,l+1}_{AB} = f_{l,l+1}\omega_0$, where $\omega_0$ is the interlayer coupling for chiral TBG. We can now rearrange the spinor to be in the sublattice basis and rescale the coordinates $r\to k_{\theta_0} r$ for some small $\theta_0$ which is commensurate with the angles $\theta_l$ \cite{TKV19},

\begin{equation}
    \mathcal{H}_{TMG} = \begin{pmatrix}
     0 & \mathcal{D}^\dagger(r) \\
     \mathcal{D}(r) & 0 
    \end{pmatrix}
\end{equation}

Where,  $\mathcal{D}$ is defined as in eq. (\ref{eq:definitionD}) and this is the Hamiltonian defined in eq. \ref{eq:Ham1}.

\section{Flat bands in chiral tri-layer systems}\label{tripod}
In this Appendix we present numerical calculations of the spatial distribution of the Dirac zero modes at the $\Gamma,K,K'$ points at the three layers of a tri-layer. We consider a chiral tri-layer system, where $\theta_{1,2}=\theta_{2,3}= 1.57^0$, with an ABA arrangement. 

\subsection{Spatial distribution of the Dirac zero modes in a tri-layer system}

\begin{figure}[t]
\begin{tabular}{c}
   \includegraphics[width=0.45\textwidth]{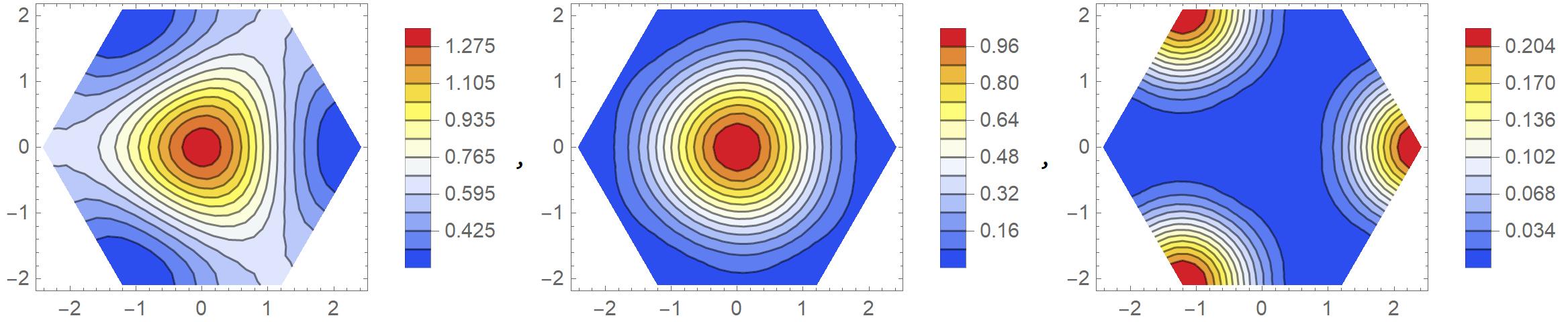} \\
   \includegraphics[width=0.45\textwidth]{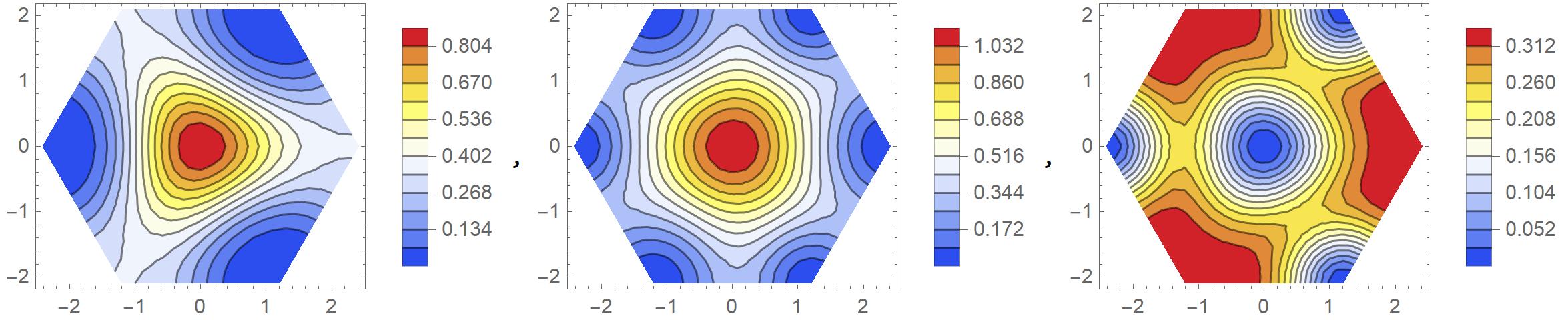}
    \end{tabular}
   \caption{Charge densities per layer for the two flat bands at the $K$ point in the Brillouin zone at the first magic angle. There are two flat bands, with Chern numbers ${\it C} = -2 , 1$. Top row: ${\it C} = -2$ band. Bottom row: ${\it C} = 1$ band. Each band is localized in a different sublattice. }
    \label{fig:densk}
\end{figure}

\begin{figure}[t]
\begin{tabular}{c}
   \includegraphics[width=0.45\textwidth]{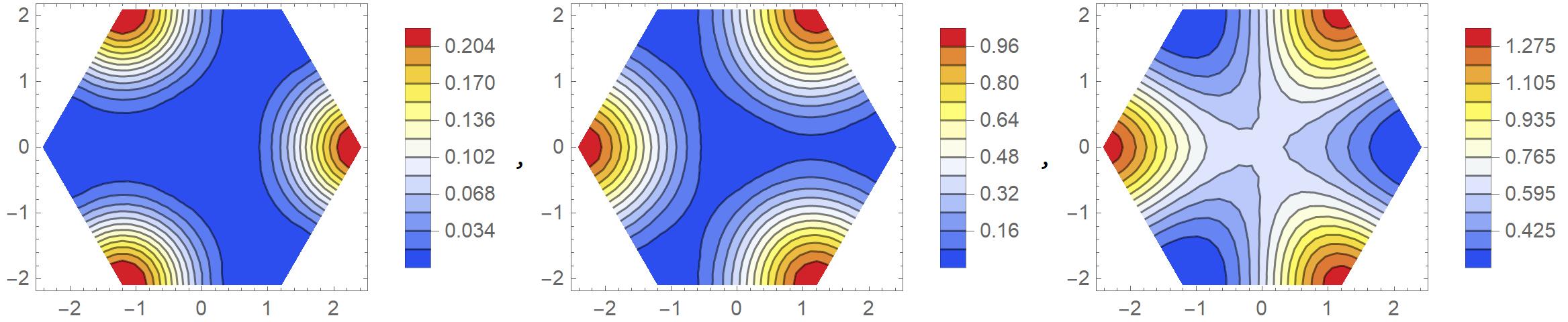} \\
   \includegraphics[width=0.45\textwidth]{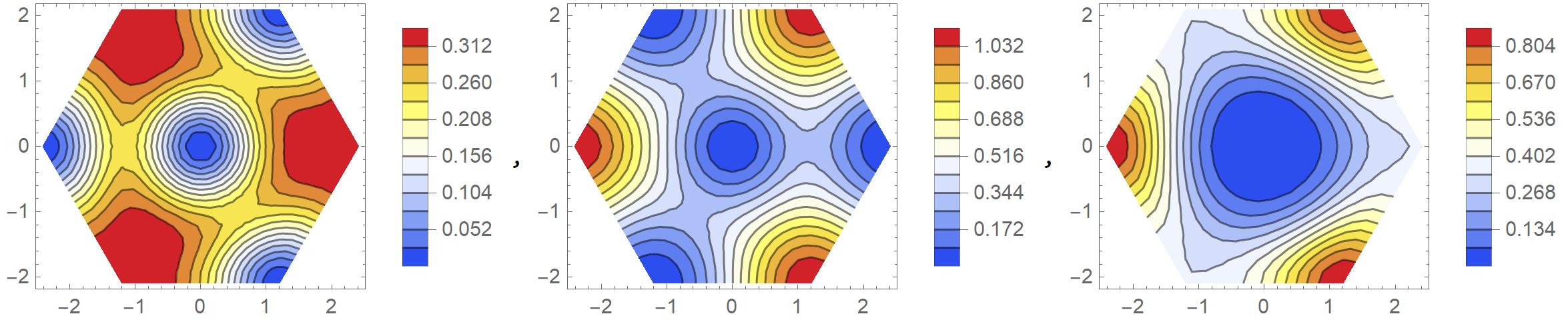}
    \end{tabular}
   \caption{As in Fig.[\ref{fig:densk}], but for point $K'$. }
   \label{fig:denskp}
\end{figure}

\begin{figure}[t]
\begin{tabular}{c}
   \includegraphics[width=0.45\textwidth]{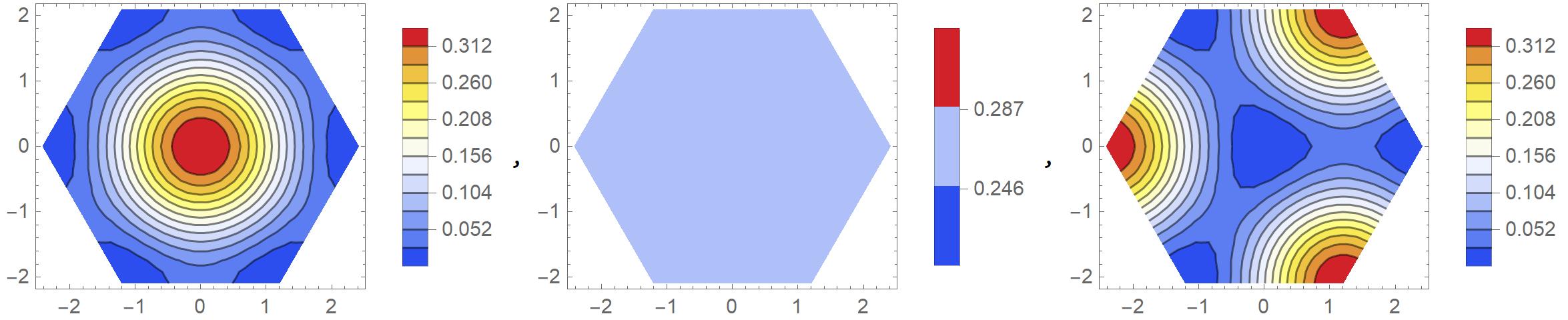} \\
   \includegraphics[width=0.45\textwidth]{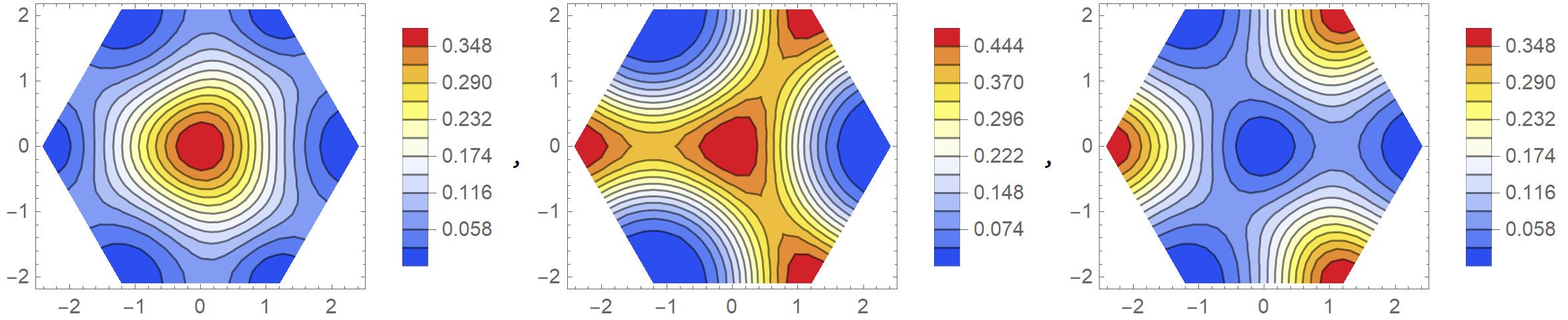}
    \end{tabular}
   \caption{As in Fig.[\ref{fig:densk}], but for point $\Gamma$. The charge density is constant at the central layer for the ${\it C} = - 2$ band, see Appendix \ref{tripod}.}
   \label{fig:densg}
\end{figure}

\begin{figure}[t]
\includegraphics[width=0.45\textwidth]{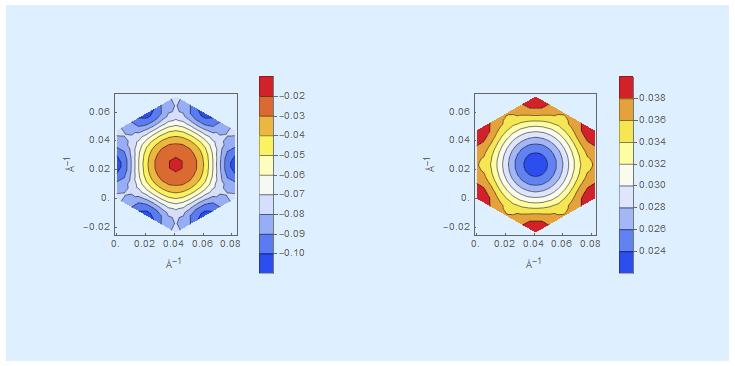}
   
   \caption{Distribution of Berry curvature for the two flat bands of an helical trilayer at the first magic angle. Left: Band with Chern number ${\it C} = -2$. Right: Band with Chern number ${\it C} = +1$.}
    \label{fig:berry}
\end{figure}

The figures below present our numerical results for the density distribution of the zero modes (Figs. \ref{fig:densk}--\ref{fig:densg}) and the distribution of the Berry curvature for the two flat bands (Fig. \ref{fig:berry}). 

The structure observed at the $\Gamma$ point, seen in Fig. (\ref{fig:densg}), may be understood using a tripod model, as we discuss in the next subsection.

\subsection{Tripod model for the wavefunction at \texorpdfstring{$\Gamma$}{Gamma}}

The continuum model for the electronic structure of an helical trilayer graphene stack is based on an expansion on plane waves, defined by the moiré periodicity\cite{LPN07}. For twisted bilayer graphene a truncation of this expansion, the so called tripod model, leads to a simple, analytical, approximation of the first magic angle\cite{BM11}. We show here that a similar approach describes {\it exactly} some states of helical trilayer graphene at the $\Gamma$ point.

We analyze helical trilayer graphene with the configuration $ABA$. This arrangement implies that top and bottom layers are rotated with respect to each other by and angle $2 \theta$ starting from an $AA$ arrangement, while the central layer is rotated with respect to the top layer by an angle $\theta$ with respect to the $AB$ stacking.

    \label{tunn}

In a momentum representation, the interlayer tunneling terms, eq.(\ref{tunn}), lead to matrix elements which combine plane waves with different momenta in different layers.

\begin{figure}[ht!]
   \includegraphics[width=0.4\textwidth]{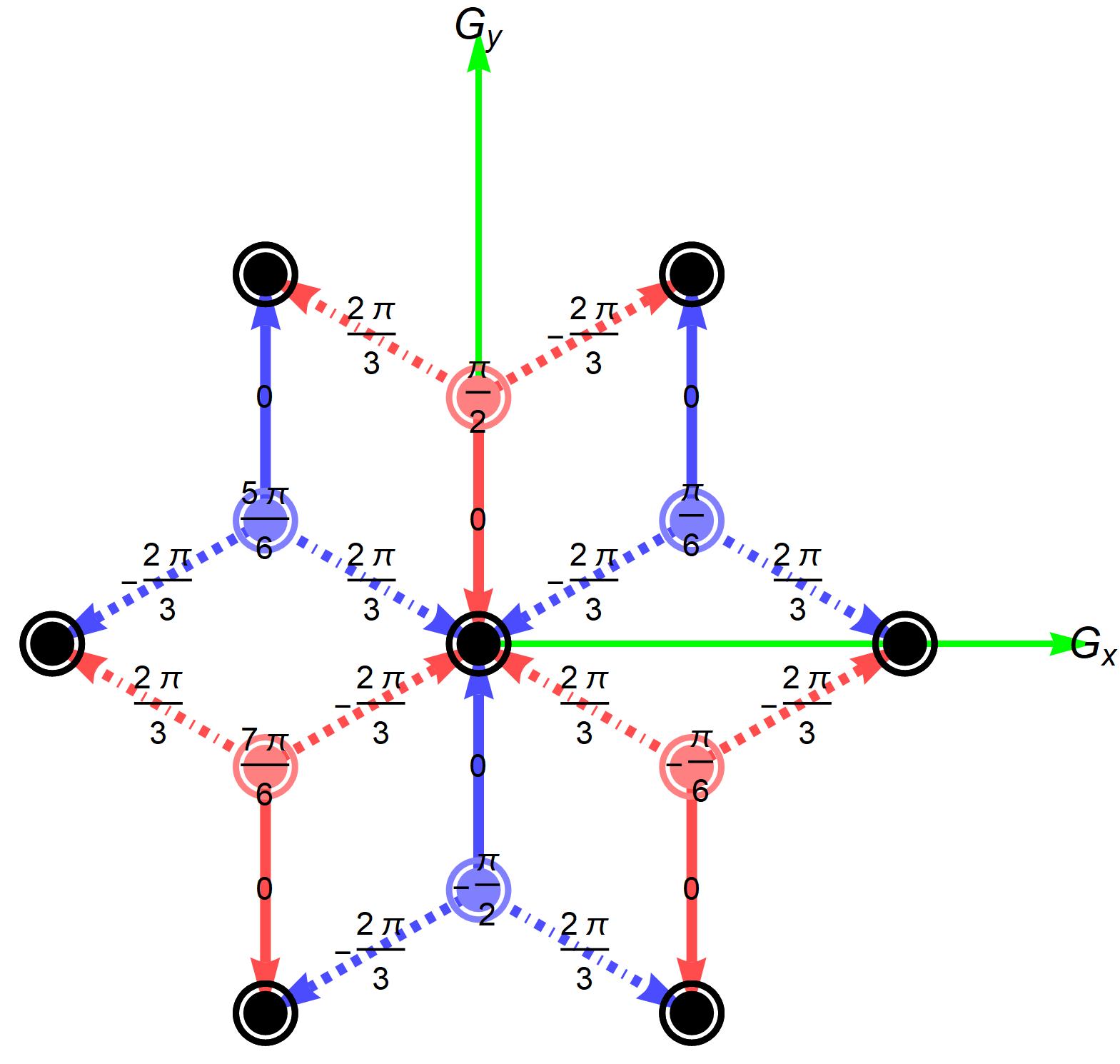} 
   \caption{Representation of the three stars of reciprocal lattice vectors used in the definition of the hamiltonian. 
   Blue points are in layer 1 (top), black points are in layer 2 (center), and red points are in layer 3 (bottom). Lines describe the connections define in eq.(\ref{tunn}). 
   For each wavevector the hamiltonian has two entries which define the $A$ and $B$ sublattices. These entries are coupled by a term $h_{AB} =  v_F K_\theta ( G_x + i G_y )$, where $K_\theta = ( 4 \pi / 3 d ) 2 \sin ( \theta / 2 )$. The numbers on each point and on each bond describe the phases of the matrix elements connecting them. These phases (defined using the arrows, and the convention $B \rightarrow A$) are independent of the numerical values of the parameters of the model.}
   \label{fig:tripod}
\end{figure}

A sketch of a few stars of reciprocal lattice vectors required for the continuum hamiltonian of a helical twisted trilayer at the $\Gamma$ point is shown in Fig.[\ref{fig:tripod}].These vectors define an effective $13 \times 13$ hamiltonian similar to the $8 \times 8$ tripod hamiltonian defined in\cite{BM11} for the $K$ point. The hamiltonian has matrix elements of absolute value $t_{AB}$ between plane waves in different sublattices ($A \leftrightarrow B$), different layers,  and different wavevectors, as shown in Fig.[\ref{fig:tripod}]. In addition, for each value of the wavevector, $v_F \vec{G}$ there are intersublattice matrix elements, of absolute value $v_F | \vec{G} |$. The phases of these matrix elements are shown in Fig.[\ref{fig:tripod}].

The ${\cal C}_3$ symmetry and the phases in Fig.[\ref{fig:tripod}] lead to the existence of a zero energy state whose wavefunction has a finite weight at the $B$ sublattice in the central wavevector, $\{ G_x , G_y \} = \{ 0 , 0 \}$, in the central layer, and also finite weights on the first star of wavevectors, which defines states in the top and bottom layers. The amplitude in the next star of wavevectors, which reside in the central layer, is zero, due to a destructive interference between the two inequivalent paths which connect a point in this star to the central point. This interference, like the phases, is independent of the value of the twist angle. As a result there is a zero energy mode with a constant charge density in the central layer, localized at the $B$ sublattice. Time reversal symmetry implies that a state at the other valley will also show a constant density localized at the $B$ sublattice of the central layer.

\section{Wronskian and Flat bands} \label{app:WronskianFlatBands}
 We now show that the vanishing of the Wronskian is a sufficient condition for $N_L < 7$ and often for any $N_L$. We start by showing that the Wronskian is constant in space using the Jacobi formula:

\begin{align}
\begin{split}
        \overline{\partial} W(r) &=   \Tr \left(\textbf{adj}\left(\Psi_{K_1} \dots \Psi_{K_n}\right) \overline{\partial} \left(\Psi_{K_1} \dots \Psi_{K_n}\right)\right) \\
        &=  \Tr \left(\textbf{adj}\left(\Psi_{K_1} \dots \Psi_{K_n}\right) \frac{1}{2i}\mathcal{P}\left(\Psi_{K_1} \dots \Psi_{K_n}\right)\right) \\ 
        &=\frac{1}{2i}\det \left(\Psi_{K_1} \dots \Psi_{K_n}\right)  \Tr \left( \mathcal{P} \right) = 0
\end{split}
\end{align}
where we can see from eq. \ref{eq:tunpot} that if $D\psi = 0 $  than $\overline{\partial} \psi = \frac{1}{2i} \mathcal{P} \psi$. According to Liouville’s theorem we can see that $W(z)$ must be a constant or have a diverging point. Since a divergence is not physical we can see that $W(r)$ must be a constant $W_0$. We show by contradiction that we can only have perfectly flat bands when the Wronskian is zero. If $W_0\ne 0 $ then the Dirac zero modes fully span the space of possible zero energy solutions and therefore any zero energy wave function can be written as,
\begin{equation}\label{eq:SumCoeff}
    \psi_k(r) = \sum_{i=1}^{N_L} c_{i}(r)\Psi_{K_i}(r)
\end{equation}
where $\Psi_{K_i}(r)$ and $\psi_k(r)$ are n-component vectors and $c_i(r)$ are scalars. 

We can now apply the operator $\mathcal{D}$ to this new zero energy wave function,
\begin{equation} \label{eq:ConstantCoefficient}
    0 = \mathcal{D}\psi_k(r) = \sum_{i=1}^{N_L} \left(\overline{\partial}c_{i}(r)\right)\Psi_{K_i}(r)
\end{equation}
as we have assumed $W_0 \ne 0$ we know that the Dirac zero modes are linearly independent and therefore we must have $\overline{\partial}c_{i}(r) = 0$, which is only possible if $c_i(r)=c_i(z)$. Then from Liouville’s theorem $c_i(z)$ must be constant or have divergences. However, since $W_0\ne 0$ we know that the sum (\ref{eq:SumCoeff}) never vanishes and $c_i(z)$ can not physically diverge. Consequently $c_i(z) = c_i$. However, a superposition of Dirac zero modes is a Dirac zero mode, and that means we can not have flat bands because the only zero energy states are at the Dirac points. 

We now show that $W_0=0$ is a sufficient condition for the existence of flat bands by showing that it implies the existence of a point $r_0$ at which a particular superposition of Dirac zero modes vanishes,
\begin{equation} \label{eq:VanishSum}
    \sum_{i=1}^m a_i\Psi_{K_i}(r_0) = 0 
\end{equation}
where $a_i$ are some constant coefficients which are not all equal to zero. This superposition can be expanded to a full flat band by multiplying it by a Jacobi theta function as previously shown in \cite{AdyYarden} \cite{TKV19}. For the specific case of Trilayer it was shown that it is possible to calculate $a_i$ from symmetry considerations \cite{GMM23}. 

We can describe every Dirac point zero energy solution $\Psi_k$ as,
\begin{equation} \label{eq:DiracModesCondition}
    \Psi_k(r) = \sum_{i=1}^m c_{i}(r)\Psi_{K_i}(r) 
\end{equation} 
where $m$ is the number of linearly independent Dirac zero modes, and $m<N_L$ since $W_0 = 0$. We will show by contradiction that  $c_i(r)$ must depend on $r$ using the existence of $N_L$ zero energy solutions at the Dirac point. If the $c_i$'s are constant then,
\begin{equation}\label{eq:ConstantC}
    \Psi_{K_{m+1}} = \sum_{i=1}^m c_{i}\Psi_{K_i}(r)
\end{equation}
To satisfy  Bloch's theorem the $c_i$'s should be non-zeros only for $K_i$'s which are identical to $K_{m+1}$.  We also require that $\Psi_{K_{m+1}}\ne \Psi_{K_i}$, and therefore we need at least two independent Dirac zero modes at this Dirac point for Eq. (\ref{eq:ConstantC}) to hold. Consequently,  as long as  $m < 6$  than $c_i$'s must depend on $r$.

If a subset of the  $c_i$'s depend on $r$ then by applying the operator $\mathcal{D}$ as in eq. (\ref{eq:ConstantCoefficient}) we find $c_{i}(r)=c_i(z)$ and by Liouville's theorem $c_i(z)$ must have  diverging points $r_{0,i}$. In order to have Eq. (\ref{eq:DiracModesCondition}) finite on both sides, there must be a vanishing superposition of Dirac zero modes satisfying eq. (\ref{eq:VanishSum}). 

We explicitly construct the flat band using Jacobi Theta functions as,

\begin{equation}
    f_k(z) = e^{iK_1 z}\frac{\vartheta_1(z - k)}{\vartheta_1(k) \vartheta_1(z)}
\end{equation}
and the flat bands as:
\begin{equation}
    \psi_k(r) = \sum_{i=1}^m a_i f_{k-K_i}(r-r_0)\Psi_{K_i}(r)
\end{equation}
We can explicitly see that the number of simple poles is equal to the number of non-zero $a_i$'s. That number is the absolute value of the band's Chern number. For two flat bands, these numbers are $n_A,n_B$ of the main text. 

\section{Chern number analysis} \label{app:ChernNumbers}
We will prove eq. (\ref{eq:ChernEquality}), first we formally define the terms $n_A$ and $n_B$ as the number of Dirac spinors used to construct the Chiral and anti-chiral flat bands, 
\begin{align}
    n_A &= \dim V_{A} = \textbf{Sp}\left(\{\Psi_i\}_{i=1}^{n_A}\right) \\
    n_B &= \dim V_{B} = \textbf{Sp}\left(\{\chi_i\}_{i=1}^{n_B}\right) \\
\end{align}
where we can see that it is equivalent to the dimension of the spinor spans $V_A$ and $V_B$, we can immediately see that:
\begin{align}
    V_{A} &\subset \ker \mathcal{D} \\
     V_{B} &\subset \ker \mathcal{D}^\dagger
\end{align}
We will show that,
\begin{align} \label{eq:FormalChernNumbers}
    N_L &= \dim(\ker \mathcal{D}^\dagger \bigoplus \ker \mathcal{D}) \\ 
    N_L &= n_D + \dim(V_A \bigoplus V_B)
\end{align}
where the second equation follows the first immediately because any Dirac spinor that is not part of a flat band construction remains protected by the original symmetries. 
We show in \ref{app:WronskianFlatBands} that when a flat band exists $m = \dim\ker \mathcal{D}$ and a set of $m$ linearly independent spinors exists that only vanish at $r_0$,
\begin{equation}
    \sum_{i=1}^m c_i \Psi_{K_i}(r_0) = 0 
\end{equation}
We can now observe that the linear dependent need not include all $m$ Dirac zero modes as such there is a minimal  value $\tilde{m}$  for whom $0<\tilde{m}<m$  and,
\begin{equation} 
    \sum_{i=1}^{\tilde{m}} c_i \Psi_{K_i}(r_0) = 0 
\end{equation}
while it is technically possible to have multiple vanishing points $r_i$ that would require multiple independent tuning parameters in order to do and generate sets of flat bands that are effectively independent of each other for this calculation despite existing at the same time. 

We can first examine the dimensionality of the anti-chiral Dirac zero modes. We can extend the $m$ Dirac spinors into a complete geometric basis in any point $r$  by introducing a new set of completing vectors such that.
\begin{equation} \label{eq:chernDet}
    \det\left(\{\psi_{K_i}\}_{i=1}^{N_L}\right) = 1
\end{equation}
where the choice of the basis completing vectors is arbitrary, we can always choose for any $r$ except $r_0$ vectors such that the value remains $1$ and the vectors are differentiable. We can now define a new alternating linear map using the potential matrix, $\mathcal{P}$ as,

\begin{align}
\begin{split}
    \sum_{t=1}^{N_L} \det \left(\left\{\mathcal{G}_{it}\psi_{K_i}\right\}_{i=1}^{N_L}\right) \\
    \mathcal{G}_{it} = \begin{cases}
        i=t &: \mathcal{P} \\
        i\ne t &: I_{N_L\times N_L} \\
    \end{cases}
\end{split}
\end{align}
where $\mathcal{G}_{it}$ is an $N_L$ by $N_L$ matrix whose values depend on $i$ and $t$, and the map is effectively the sum of $N_L$ determinants each of an $N_L \times N_L$ matrix. From the uniqueness of alternating linear maps we can immediately see that,
\begin{equation}
    \sum_{t=1}^{N_L} \det\left(\{\mathcal{G}_{it}\psi_{K_i}\}_{i=1}^{N_L}\right) = \Tr{\mathcal{P}} \det\left(\{\psi_{K_i}\}_{i=1}^{N_L}\right) = 0
\end{equation}
where we know from uniqueness that the new map will be scalar times the determinant. We can see that the the scalar would be the trace of $\mathcal{P}$ by considering the map for the standard vector basis. 

We can use this result to show eq. (\ref{eq:FormalChernNumbers}) by showing that  $\left( \ker \overline{\mathcal{D}}\right)^\perp \subset \ker D^\dagger$   and the reverse is also true $\left( \ker \mathcal{D}^\dagger\right)^\perp \subset \ker \overline{D} $.  Therefore  $\dim(\ker \mathcal{D}^\dagger \bigoplus \ker \mathcal{D} ) = N_L$ which is the required result. We start by differentiating the determinant in eq. (\ref{eq:chernDet}), 
\begin{align}
\begin{split}
   0&= \overline{\partial} \det \left(\{\psi_{K_i}\}_{i=1}^{N_L}\right)\\ 
   &= \sum_{t=1}^{m} \det\left(\{{\mathcal{G}_{it}}{\psi_{K_i}}\}_{i=1}^{N_L}\right) \\
   &+  \sum_{t=m+1}^{N_L} \det\left(\{\mathcal{R}_{it}\psi_{K_i}\}_{i=1}^{N_L}\right) \\
   &= - \sum_{t=m+1}^{N_L} \det\left(\{\mathcal{G}_{it}\psi_{K_i}\}_{i=1}^{N_L}\right) \\
   &+  \sum_{t=m+1}^{N_L} \det \left(\{\mathcal{R}_{it}\psi_{K_i}\}_{i=1}^{N_L}\right) \\
   & = \sum_{t=m+1}^{N_L} \det\left(\{\mathcal{D}_{it}\psi_{K_i}\}_{i=1}^{N_L}\right)\\
   &=\sum_{t=m+1}^{N_L} C_t \mathcal{D} \psi_t\\
    \mathcal{R}_{it} &= \begin{cases}
        i=t &: \overline{\partial} \\
        i\ne t &: I_{N_L\times N_L}
    \end{cases} \\ 
     \mathcal{D}_{it} &= \begin{cases}
        i=t &: \mathcal{D} \\
        i\ne t &: I_{N_L\times N_L}
    \end{cases}
\end{split}
\end{align}
where $C_t$ is the cofactor $t$ column.  We know that $\forall t > m: \; \mathcal{D}_{it}\psi_{i} \ne 0$  and that we have no repeat values. We know $C_t \parallel \psi_t$  and we can write the complex conjugation of the above as:
\begin{align}
\begin{split}
0 = \sum_{t=m+1}^{N_L} l_t (r) \overline{\psi_t} \mathcal{D}^\dagger \overline{\psi_t}\\
\end{split}
\end{align}
where $l_t$ are some space dependent periodic coefficients. We can see the expression does not depend on the choice of $\psi_t$ or $r$ and therefore $\psi_t \in \ker \mathcal{D}^\dagger \bigoplus \ker \mathcal{D} $. Furthermore, we know that $\psi_t \perp \ker \mathcal{D}$ and as such $\psi_t \in \ker \mathcal{D}^\dagger $ and that proves that t$\left( \ker \overline{\mathcal{D}}\right)^\perp \subset \ker D^\dagger$. We can see from this result or from \ref{app:WronskianFlatBands} that we will have an anti-Chiral solution, 
\begin{equation}
    \sum_{i=1}^{n_B} c_i\chi_{K_i}(r_0) = 0
\end{equation}
where $n_B \le N_L - n_A$ and $\chi_{K_i}$ are the other Dirac spinors. We can again see that the solution need not depend on all Dirac zero modes and as such $n_B <\le \dim \ker \mathcal{D}^\dagger$.  

The remaining question is does a non-trivial intersection exist between the two kernels, we can see that this is possible for the case with no shifts and as an intersection term represents a repeating zero mode.

 We can directly infer $n_A$ and $n_B$ from the flat band curves for curves of types $1-2$  but not for curves of type $3-4$ whose Dirac zero modes aren't fully determined. Furthermore, each of these sets may contribute $1$  or $2$ flat bands depending on the order of the zeros. A few examples of the calculation for cases we observe for $N_L=3,5$  are seen in Table \ref{tab:ChernNumbers}. We can calculate $N_L$ and $n_D$ numerically for TMG as seen in Table \ref{tab:multilayernumerics} which we can use to understand the possible Chern behavior without complete calculation of the Berry curvature. 
 We can directly infer $n_A$ and $n_B$ from the flat band curves for curves of types $1-2$  but not for curves of type $3-4$ whose Dirac zero modes aren't fully determined. Furthermore, each of these sets may contribute $1$  or $2$ flat bands depending on the order of the zeros. A few examples of the calculation for cases we observe for $N_L=3,5$  are seen in Table \ref{tab:ChernNumbers}. We can calculate $N_L$ and $n_D$ numerically for TMG as seen in Table \ref{tab:multilayernumerics} which we can use to understand the possible Chern behavior without complete calculation of the Berry curvature. 
 
 \begin{widetext}
\begingroup

\setlength{\tabcolsep}{10pt} 
\renewcommand{\arraystretch}{1.3} 
  \begin{table*}[!ht]
\caption{Possible combinations of Chern numbers by type and the number of layers $N_L$. Here $|C_A|$ and $|C_B|$ are the absolute values of the Chern numbers per band.}
 \label{tab:ChernNumbers} \begin{tabular}{||c|c|c|c|c|c|c||}
   \hline\hline
    Type & shift &\# zero bands & Parity &  $|C_A|$ & $|C_B|$ & $n_D$ \\ \hline
    1 &  No&4&Either& $1$ or $2$& $1$ or $2$& $N_L - 2$ ,   $N_L - 3$ ,  $N_L- 4$ \\ \hline
 2\
  &  No & 2 & Even & 1 & 1 &$N_L -2 $ \\\hline
    2 & Yes &2 & Either & $2$ & $1$ & $N_L - 3$  \\ 
    \hline
    3 & Yes &4 &Either & $1$ & $1$  & $N_L - 2$\\\hline
 3 & Yes & 4 & Either & 2 & 2&$N_L -4$\\
    \hline
    4 & Yes &4 &Either & $2$ & $1$ & $N_L - 3$\\
    \hline\hline\end{tabular}
\end{table*}  
\endgroup
\end{widetext}

\begingroup
\setlength{\tabcolsep}{11pt}
\renewcommand{\arraystretch}{1.5}

\begin{center}
\begin{table*}[!tbp]
          \caption{Magic angles by N-layer stacks where next nearest neighbors are rotated by the same angle in the same direction. Results are shown for no interlayer shift, and for an interlayer shift at the center of the stack equal to the distance between the center and the corner of the unit cell of monolayer graphene.} \label{tab:multilayernumerics}
          \begin{tabular}{||c|c|c|c|c|c|c||}
         \hline
         \hline
       $N_L$ & arrangement &angle &velocities &vanishing &\#  &
       $n_D$  \\
        & & & $\{ K , K' , \Gamma \}$ &velocities &zero  &
        \\
        & & &  & & bands &
       \\
       \hline\hline
        3 &AAA &$0.715^\circ$ &$\{ \{v_1 \} , \{ v_1 \} , \{v_2 \} \}$ &$v_1 , v_2 = 0$ &4 &1 \\
        \hline
       3 &ABA &$0.34^\circ$ &$\{ \{v_1 \} , \{ v_1 \} , \{v_2 \} \}$ &$v_1 , v_2 = 0$ &2 &0 \\
       \hline
       3 &ABA &$0.495^\circ$ &$\{ \{v_1 \} , \{ v_1 \} , \{v_2 \} \}$ &$v_1 = 0$ &4 &1 \\
       \hline
       3 &ABA &$1.57^\circ$ &$\{ \{v_1 \} , \{ v_1 \} , \{v_2 \} \}$ &$v_1 , v_2 = 0$ &2 &0 \\
       \hline \hline
       4 &AAAA &$0.55^\circ$ &$\{ \{v_1 , v_1 \} , \{ v_2 \} ,\{v_2 \} \}$ &$v_2 = 0$ &2 &2\\
       \hline
       4 &AAAA &$0.8637^\circ$ &$\{ \{v_1 , v_1 \} , \{ v_2 \} ,\{v_2 \} \}$ &$v_1, v_2 = 0$ &4 &0\\
        \hline
       4 &BAAB &$0.46^\circ$ &$\{ \{v_1 , v_2 \} , \{ v_3 \} \{v_3 \} \}$ &$v_1, v_3 = 0$ &2 &1\\
        \hline
       4 &BAAB &$0.64^\circ$ &$\{ \{v_1 , v_2 \} , \{ v_3 \} , \{v_3 \} \}$ &$v_1, v_3 = 0$ &4 &1\\
        \hline
       4 &BAAB &$1.84^\circ$ &$\{ \{v_1 , v_2 \} , \{ v_3 \} , \{v_3 \} \}$ &$v_1, v_3 = 0$ &2 &1\\
       \hline \hline
       5 &AAAAA &$0.55^\circ$ &$\{ \{v_1 , v_2 \}, \{v_3 \} , \{ v_1 , v_2 \}  \}$ & $v_1, v_2 = 0$ &4 &1 \\
       \hline
       5 &AAAAA &$0.85495^\circ$ &$\{ \{v_1 , v_2 \}, \{v_3 \} , \{ v_1 , v_2 \}  \}$ & $v_1, v_2 = 0$ &4 &1
       \\
        \hline
       5 &BBABB &$0.3^\circ$ &$\{ \{v_1 , v_2 \}, \{v_3 \} , \{ v_1 , v_2 \}  \}$ & $v_1 = 0$ &2 &3
       \\
       \hline
       5 &BBABB&$0.72^\circ$ &$\{ \{v_1 , v_2 \}, \{v_3 \} , \{ v_1 , v_2 \}  \}$ & $v_1 , v_3 = 0$ &4 &2
       \\
       \hline
       5 &BBABB &$0.905^\circ$ &$\{ \{v_1 , v_2 \}, \{v_3 \} , \{ v_1 , v_2 \}  \}$ & $v_1 = 0$ &2 &3
       \\
       \hline
       5 &BBABB &$1.99^\circ$ &$\{ \{v_1 , v_2 \}, \{v_3 \} , \{ v_1 , v_2 \}  \}$ & $v_1, v_3 = 0$ &2 &2
       \\
       \hline \hline
       6 &AAAAAA &$0.2253^\circ$ &$\{ \{v_1 , v_2 \} , \{ v_1 , v_2 \} , \{v_3 , v_3 \} \}$ & $v_1 = 0$ &2 &4
       \\
       \hline
        6 &AAAAAA &$0.27^\circ$ &$\{ \{v_1 , v_2 \} , \{ v_1 , v_2 \} , \{v_3 , v_3 \} \}$ & $v_1 = 0$ &4 &2
       \\
       \hline
       6 &AAAAAA &$0.326^\circ$ &$\{ \{v_1 , v_2 \} , \{ v_1 , v_2 \} , \{v_3 , v_3 \} \}$ & $v_1, v_3 = 0$ &2 &4
       \\
       \hline
       6 &AAAAAA &$0.6391^\circ$ &$\{ \{v_1 , v_2 \} , \{ v_1 , v_2 \} , \{v_3 , v_3 \} \}$ & $v_1 = 0$ &2 &4
       \\
        \hline
       6 &AAAAAA &$0.85575^\circ$ &$\{ \{v_1 , v_2 \} , \{ v_1 , v_2 \} , \{v_3 , v_3 \} \}$ & $v_1 = 0$ &4 &4
       \\
       \hline
       6 &BBAABB &$0.23^\circ$ &$\{ \{v_1 , v_2 \} , \{ v_1 , v_2 \} , \{v_3 , v_4 \} \}$ & $v_1, v_3 = 0$ &4 &3
       \\
        \hline
       6 &BBAABB &$0.32^\circ$ &$\{ \{v_1 , v_2 \} , \{ v_1 , v_2 \} , \{v_3 , v_4 \} \}$ & $v_1, v_3 = 0$ &2 &3
       \\
        \hline
       6 &BBAABB &$0.35^\circ$ &$\{ \{v_1 , v_2 \} , \{ v_1 , v_2 \} , \{v_3 , v_4 \} \}$ & $v_1, v_3 = 0$ &2 &3
       \\
       \hline
       6 &BBAABB &$0.77^\circ$ &$\{ \{v_1 , v_2 \} , \{ v_1 , v_2 \} , \{v_3 , v_4 \} \}$ & $v_1, v_3 = 0$ &4 &3
       \\
       \hline
       6 &BBAABB &$1.295^\circ$ &$\{ \{v_1 , v_2 \} , \{ v_1 , v_2 \} , \{v_3 , v_4 \} \}$ & $v_1, v_3 = 0$ &2 &3
       \\
       \hline
       6 &BBAABB &$2.08^\circ$ &$\{ \{v_1 , v_2 \} , \{ v_1 , v_2 \} , \{v_3 , v_4 \} \}$ & $v_1, v_3 = 0$ &2 &2
       \\ \hline \hline
        7 &AAAAAAA &$0.572401^\circ$ &$\{ \{v_3 , v_4 , v_5 \} , \{v_1 , v_2 \} , \{ v_1 , v_2  \}  \}$ & $v_1 , v_3 = 0$ &4 &4
       \\
        \hline 
       7 &AAAAAAA &$0.856^\circ$ &$\{ \{v_3 , v_4 , v_5 \}  ,\{v_1 , v_2 \} , \{ v_1 , v_2  \}  \}$ & $v_1 , v_3 = 0$ &4 &4
       \\
        \hline 
       7 &ABBABBA &$0.23^\circ$ &$\{ \{v_3 , v_4 , v_5 \} ,\{v_1 , v_2 \} , \{ v_1 , v_2  \} $ & $v_1 , v_3 = 0$ &4 &4
       \\
        \hline 
       7 &ABBABBA &$0.298^\circ$ &$\{ \{v_3 , v_4 , v_5 \} ,\{v_1 , v_2 \} , \{ v_1 , v_2  \} $ & $v_1 , v_3 = 0$ &4 &4
       \\
        \hline 
       7 &ABBABBA &$0.306^\circ$ &$\{ \{v_3 , v_4 , v_5 \} ,\{v_1 , v_2 \} , \{ v_1 , v_2  \} $ & $v_1 , v_3 = 0$ &2 &4
       \\
        \hline 
       7 &ABBABBA &$0.4^\circ$ &$\{ \{v_3 , v_4 , v_5 \} ,\{v_1 , v_2 \} , \{ v_1 , v_2  \} $ & $v_1 , v_3 = 0$ &2 &4
       \\
        \hline 
       7 &ABBABBA &$0.44^\circ$ &$\{ \{v_3 , v_4 , v_5 \} ,\{v_1 , v_2 \} , \{ v_1 , v_2  \} $ & $v_1 , v_3 = 0$ &4 &4
       \\
        \hline 
         \multicolumn{7}{||r||}{{Continued on the next page }} \\ \hline \hline
\end{tabular}
\end{table*}
\end{center}
\endgroup
\newpage
\begingroup
\setlength{\tabcolsep}{8pt}
\renewcommand{\arraystretch}{1.5}
\begin{center}
\begin{table*}[!htb]
     \label{tab:multilayernumerics2}
          \begin{tabular}{||c|c|c|c|c|c|c||}
         \hline
         \hline
        $N_L$ & arrangement &angle &velocities &vanishing &\#  &
       $n_D$ \\
        & & & $\{ K , K' , \Gamma \}$ &velocities &zero  &
        \\
        & & &  & & bands &
       \\
        \hline \hline
               7 &ABBABBA &$0.8^\circ$ &$\{ \{v_3 , v_4 , v_5 \} ,\{v_1 , v_2 \} , \{ v_1 , v_2  \} $ & $v_1 , v_3 = 0$ &4 &4
       \\
        \hline 
       7 &ABBABBA &$2.14^\circ$ &$\{ \{v_3 , v_4 , v_5 \} ,\{v_1 , v_2 \} , \{ v_1 , v_2  \} $ & $v_1 , v_3 = 0$ &2 &4
       \\ \hline \hline
       8 &AAAAAAAA &$0.404^\circ$ &$\{ \{v_1 , v_2 , v_3 \} ,\{v_4 , v_4 \} , \{ v_1 , v_2 , v_3 \} $ & $ v_1 = 0$ &2 &6
       \\ \hline
       8 &AAAAAAAA &$0.59^\circ$ &$\{ \{v_1 , v_2 , v_3 \} ,\{v_4 , v_4 \} , \{ v_1 , v_2 , v_3 \} $ & $ v_1 , v_4 = 0$ &4 &4
       \\
       \hline
       8 &AAAAAAAA &$0.856^\circ$ &$\{ \{v_1 , v_2 , v_3 \} ,\{v_4 , v_4 \} , \{ v_1 , v_2 , v_3 \} $ & $ v_1 , v_2 = 0$ &4 &4
       \\
         \hline
       8 &ABBAABBA &$2.18^\circ$ &$\{ \{v_1 , v_2 , v_3 \} ,\{v_4 , v_5 \} , \{ v_1 , v_2 , v_3 \} $ & $ v_1 , v_4 = 0$ &2 &5
       \\ \hline \hline
        9 &AAAAAAAAA &$0.23^\circ$ &$\{ \{v_1 , v_2 , v_3 \} , \{ v_1 , v_2  , v_3 \} , \{v_4 , v_5 , v_6 \} \}$ & $v_1 , v_2  = 0$ &4 &5
       \\
       \hline
       9 &AAAAAAAAA &$0.266^\circ$ &$\{ \{v_1 , v_2 , v_3 \} , \{ v_1 , v_2  , v_3 \} , \{v_4 , v_5 , v_6 \} \}$ & $v_1 , v_4  = 0$ &4 &6
       \\
       \hline
       9 &AAAAAAAAA &$0.342^\circ$ &$\{ \{v_1 , v_2 , v_3 \} , \{ v_1 , v_2  , v_3 \} , \{v_4 , v_5 , v_6 \} \}$ & $ v_1 , v_2  = 0$ &4 &5
       \\
       \hline
       9 &AAAAAAAAA &$0.418^\circ$ &$\{ \{v_1 , v_2 , v_3 \} , \{ v_1 , v_2  , v_3 \} , \{v_4 , v_5 , v_6 \} \}$ & $v_1 , v_4 , v_5 = 0$ &4 &5
       \\
       \hline
        9 &AAAAAAAAA &$0.5875^\circ$ &$\{ \{v_1 , v_2 , v_3 \} , \{ v_1 , v_2  , v_3 \} , \{v_4 , v_5 , v_6 \} \}$ & $v_1 , v_2 = 0$ &4 &5
       \\
        \hline
       9 &AAAAAAAAA &$0.85575^\circ$ &$\{ \{v_1 , v_2 , v_3 \} , \{ v_1 , v_2  , v_3 \} , \{v_4 , v_5 , v_6 \} \}$ & $v_1 , v_4 , v_5 = 0$ &4 &5
       \\
              \hline
        9 &BABBABBAB &$0.30^\circ$ &$\{ \{v_1 , v_2 , v_3 \} , \{ v_1 , v_2  , v_3 \} , \{v_4 , v_5 , v_6 \} \}$ & $v_1 , v_4 = 0$ &4 &6
       \\
         \hline
        9 &BABBABBAB &$0.335^\circ$ &$\{ \{v_1 , v_2 , v_3 \} , \{ v_1 , v_2  , v_3 \} , \{v_4 , v_5 , v_6 \} \}$ & $v_1 , v_4 = 0$ &4 &6
       \\
         \hline
        9 &BABBABBAB &$0.355^\circ$ &$\{ \{v_1 , v_2 , v_3 \} , \{ v_1 , v_2  , v_3 \} , \{v_4 , v_5 , v_6 \} \}$ & $v_1 , v_4 = 0$ &2 &6
       \\
        \hline
        9 &BABBABBAB &$0.63^\circ$ &$\{ \{v_1 , v_2 , v_3 \} , \{ v_1 , v_2  , v_3 \} , \{v_4 , v_5 , v_6 \} \}$ & $v_1 , v_4 = 0$ &2 &6
       \\
         \hline
        9 &BABBABBAB &$0.84^\circ$ &$\{ \{v_1 , v_2 , v_3 \} , \{ v_1 , v_2  , v_3 \} , \{v_4 , v_5 , v_6 \} \}$ & $v_1 , v_4 = 0$ &4 &6
       \\
        \hline
        9 &BABBABBAB&$1.16^\circ$ &$\{ \{v_1 , v_2 , v_3 \} , \{ v_1 , v_2  , v_3 \} , \{v_4 , v_5 , v_6 \} \}$ & $v_1 , v_4 = 0$ &2 &6
       \\
        \hline
        9 &BABBABBAB &$1.825^\circ$ &$\{ \{v_1 , v_2 , v_3 \} , \{ v_1 , v_2  , v_3 \} , \{v_4 , v_5 , v_6 \} \}$ & $v_1 , v_4 = 0$ &2 &6
       \\
       \hline
        9 &BABBABBAB &$2.212^\circ$ &$\{ \{v_1 , v_2 , v_3 \} , \{ v_1 , v_2  , v_3 \} , \{v_4 , v_5 , v_6 \} \}$ & $v_1  , v_4 = 0$ &2 &6  \\ 
       \hline
       \hline
\end{tabular}
\end{table*}
\end{center}

\endgroup

\section{Mapping of the flat bands wave functions onto lowest Landau level wave functions}\label{app:LandauLevelComparison}
We will now show that a flat band with a Chern number $N$ will be spanned by $N$ lowest Landau Levels each with a density of states $1/N$ compared to the original flat band, and each multiplied by a different periodic function $G^l$. We examine flat bands constructed using the vanishing of the zero modes as can be seen in Appendix \ref{app:WronskianFlatBands}.

\begin{equation} \label{eq:ll:superposition}
    \sum_{i=1}^N c_i \Psi_{K_i}(r_0) = 0
\end{equation}

In order to construct a periodic solution we must find quasi-periodic functions $h_i$ that maintain this relationships such that, 
\begin{equation}
    \sum_{i=1}^N c_i h_i(r_0 + na_1 + ma_2)\Psi_{K_i}(r_0 + n a_1 + m a_2) = 0
\end{equation}
where $a_{1,2}$ are the real space periods of the problem. To do so we define the following function $N$ functions,

\begin{equation}
    h_{K_i}^l(z) =e^{iK_iz} \sum_{t=1}^{N} e^{2\pi i \alpha l t}\vartheta \left(\alpha (z+ t)  - K_i|\alpha \omega \right)
\end{equation}
where $\alpha = \frac{N-1}{N}$, these functions maintain periodic relations
\begin{align}
    \left|\frac{h_{K_i}^l(z + a_1)}{h_{K_i}^{l}(z)}\right| &= 1 \\
    \left|\frac{h_{K_i}^{l}(z + a_2)}{h_{K_i}^{l+1}(z)}\right| &= e^{-\frac{\pi \alpha (\Im a_2)^2}{\Im \omega}} 
\end{align}

These functions are orthogonal and cancel the phase that the zero mode solutions gain, we can now define the new wave function solutions,
\begin{align}
    \psi^l(r) = \frac{ \sum_{i=1}^N c_i h_{K_i}^{l}(z-z_0)\Psi_{K_i}(z-z_0)}{\vartheta_1(z-z_0|\omega)}
\end{align}
where $\psi^l$ is a function at one point in $K$ space that is decaying in $\Im z$   and allows us to construct a normalizable solution. $\psi^l$ is periodic along $a_1$ such that  $|\psi^l(z+a_1)|=|\psi^l(z)|$   and rotating along $a_2$  such that $|\psi^l(z+a_2)|=e^{- (\Im z)^2 /( N\Im \omega)}|\psi^{l+1}(z)|$. These function decay along the imaginary axis as $e^{- (\Im z)^2 /( N\Im \omega)}$.  
Both functions maintain the condition that  all the zeroes of $\vartheta_1$ are cancelled by zeroes of the sum, that  would be possible as long as the matrix $\left(f_i^l\right)_{il}$  has a determinant of 0 at the vanishing points $r_0 + ma_1 + na_2$.   We can note that if we have two identical $K$ points than the matrix will have two identical columns and such a solution $\{c_i\}_{k=1}^N$ would exist that satisfies eq. (\ref{eq:ll:superposition}) and eq. (\ref{eq:ll:periodic}). We note that for the $N=2$ case the $f_i^l$  are all even and as such having $\pm K$ is sufficient to create two identical columns and for $N\ge 4$ there are only 3 possible $K$ values and we must have two identical columns.   We can know define $N$ periodic G functions,

\begin{align} \label{eq:ll:periodic}
\begin{split}
    G^{l}(r) &=  e^{\frac{\pi}{2NA_m}|z-2z_0|^2}\frac{e^{-\frac{\pi}{2NA_m}(z-2z_0)^2}}{\vartheta_1(z - z_0|\omega)}\\
    & \times \sum_{i=1}^N c_i h_{K_i}^{l}(z-z_0)\Psi_{K_i}(z)
\end{split}
\end{align}
Where we maintain periodicity as follows:
\begin{align}
    G^l(r + a_1) &= U_1 G^l(r) \\ 
    G^l(r + a_2) &= U_2 G^{l+1}(r)
\end{align}
where $U_{1,2}$ are phase vectors matching the required boundary conditions. We can define $l$  normalizable wave functions that solve the original system,
\begin{equation}
     \Psi^l_{0} = e^{-\frac{\pi}{2NA_m}|z-2z_0|^2}G^{l}(r) 
\end{equation}
\begin{align}   
    \begin{split}
        \Psi^l_{b} &= e^{-\left(\frac{b}{4} + \frac{\pi}{2NA_m}\right)|z-2z_0|^2}G^{l}(r) \\
        &\equiv e^{-\frac{b_{\textit{eff}}}{4}|z-2z_0|^2}G^{l}(r) 
    \end{split}
\end{align}
where $b_{\textit{eff}} = b + \frac{2\pi}{N A_m}$ and as such we can see that the wave function would not be normalizable for $b=-\frac{2\pi}{N A_m}$ which is consistent with $b=\frac{\phi_0}{N}$ we received from the initial understanding of the system.  The system now consists of $N$ wave functions that decay at rate of $1/N$ compared to the original system.

\FloatBarrier

\bibliography{main}

\end{document}